**Statistically enhanced self-attraction of random patterns**

D. B. Lukatsky, K.B. Zeldovich, and E. I. Shakhnovich
*Department of Chemistry and Chemical Biology, Harvard University, Cambridge, MA 02138*

**Abstract**

In this work we develop a theory of interaction of randomly patterned surfaces as a generic prototype model of protein-protein interactions. The theory predicts that pairs of randomly superimposed identical (homodimeric) random patterns have always twice as large magnitude of the energy fluctuations with respect to their mutual orientation, as compared with pairs of different (heterodimeric) random patterns. The amplitude of the energy fluctuations is proportional to the square of the average pattern density, to the square of the amplitude of the potential and its characteristic length, and scales linearly with the area of surfaces. The greater dispersion of interaction energies in the ensemble of homodimers implies that strongly attractive complexes of random surfaces are much more likely to be homodimers, rather than heterodimers. Our findings suggest a plausible physical reason for the anomalously high fraction of homodimers observed in real protein interaction networks.

PACS numbers:

Interaction of disordered surface patterns is a widespread phenomenon in biological, and engineered soft and biomaterial interface systems. The general problem of interaction of random patterns represents a first step towards the fundamental understanding of the design principles of bio-molecular recognition from the first principles, using the bottom-up approach [1]. One such design principle was recently predicted in [2]. It was predicted computationally [2] that the attractive interaction in pairs of identical random surfaces (we term such self-interacting pairs of surfaces "homodimers") is statistically stronger than the attraction in pairs of different random surfaces of the same size (we term such pairs "heterodimers"). By changing the mutual orientation of the surfaces and looking for the lowest possible energy in a given pair of surfaces, it was found that in homodimers the average minimum energy of interaction between the surfaces is lower than in heterodimers [2]. The probability distribution of the minimum energies of interaction in a pair of surfaces is a type I (Gumbel) extreme value distribution (EVD).

Here, we propose a theory that confirms the universal nature of the effect observed in [2], and shows that the energy fluctuations (with respect to mutual orientation of the surfaces) in homodimers are exactly twice larger than in heterodimers. This implies that the distribution of the lowest interaction energies for homodimers is always shifted towards the lower energies as compared with heterodimers. We provide the explicit expressions for the energy fluctuations in both cases, and an estimate for the average difference in the minimum energy between homo- and heterodimers.

Our model consists of 2D disk-like flat surfaces of radius $R$ and surface area $A$, whereupon $N_0$ particles are placed; $N_0$ is assumed to be fixed (see Fig. 1). The average density of particles is $\phi_0 = N_0/A$, and for each surface the pattern is quenched, so the particles are not allowed to move. The particles interact with each other via a finite-range potential of magnitude $U_0$ and range $\xi$. To find the interaction energy $E$ in a pair of two different surfaces (heterodimer), we superimpose the surfaces in a coplanar and coaxial configuration at a separation $h$ between the surfaces, $h<\xi$. The inter-surface interaction energy $E$ is then computed for every pair of surfaces. The model of homodimers is analogous, except for the fact that in this case we superimpose each surface with the *reflected* image of itself. This superimposition represents the correct analogy of real protein surfaces forming a homodimeric interface. The probability distributions $P^{(homo)}(E)$ or $P^{(hetero)}(E)$ for homo- and heterodimers, respectively, will be derived.

For a given pair of surface patterns, the interaction energy $E$ depends on the relative rotation of the disks about their common axis, and the distribution of the energies within a given pair of disks also follows $P^{(homo)}(E)$ or $P^{(hetero)}(E)$. An orientation of the patterns corresponding to the lowest energy roughly mimics the native state of a protein complex. Therefore, the minimum energy of interaction $E_{min}$ that can be achieved in a given pair of patterns and its distributions $P_{min}^{(homo)}(E)$ and $P_{min}^{(hetero)}(E)$ are of a particular interest. The distributions $P_{min}^{(homo)}(E)$ and $P_{min}^{(hetero)}(E)$ are obtained from $P^{(homo)}(E)$ or $P^{(hetero)}(E)$ as the statistics of extreme order (extreme value distributions) of length $M \sim 2\pi R/\xi$, as within each pair there are about $M$ mutual orientations of the disks with statistically independent values of energy.

We begin by calculating the properties of $P^{(homo)}(E)$ and $P^{(hetero)}(E)$ for heterodimers and homodimers, from which we later derive the exteme value distributions $P_{min}^{(homo)}(E)$ and $P_{min}^{(hetero)}(E)$.

The energy of interaction between two surfaces in a heterodimer reads

**Eq. 1**
$$E = \int \varphi_1(\vec{\rho}_1) U(\vec{r}_1 - \vec{r}_2) \varphi_2(\vec{\rho}_2) \, d^2\vec{\rho}_1 \, d^2\vec{\rho}_2 \,,$$

where $\varphi_1(\vec{\rho}_1)$ and $\varphi_2(\vec{\rho}_2)$ are the surface densities of particles constituting the random patterns on the first and second surface, respectively. The radius-vectors are $\vec{r}_1 = (\vec{\rho}_1, -h/2)$, and $\vec{r}_2 = (\vec{\rho}_2, +h/2)$, with $\vec{\rho}_1 = (x_1, y_1)$, and $\vec{\rho}_2 = (x_2, y_2)$ being the 2D vectors on the first and second surface, respectively. The integration in Eq. 1 is performed with respect to the surface areas of both disks. The inter-particle interaction potential, $U(\vec{r}_1 - \vec{r}_2)$, between two particles located at the point $\vec{r}_1$ on the first surface and $\vec{r}_2$ on the second surface is assumed to be pairwise additive, and it depends only on the distance between the particles $|\vec{r}_1 - \vec{r}_2|$. At this point we do not make any other assumptions about $U(r)$.

The surface densities $\varphi_i(\vec{\rho})$ for each surface $i = 1, 2$ can be represented as

**Eq. 2**
$$\varphi_i(\vec{\rho}) = \phi_0 + \phi_i(\vec{\rho}),$$

where $\phi_0$, is the average surface density, and $\phi_i(\vec{\rho})$, is the deviation (or fluctuation) of the local density from its average value $\phi_0$ at a given realization of random pattern on surface $i$. Substituting Eq. 1 into Eq. 2, the energy $E$ can be represented as a sum, $E = E_0 + E_1 + E_2$. Here, $E_0$ is the average interaction energy, independent on the density fluctuations; the next contribution, $E_1$, is linear in the density fluctuations, $\phi_i(\vec{\rho})$, and the last contribution, $E_2$, is quadratic in the density fluctuations, $\phi_i(\vec{\rho})$. Our objective is to find the average fluctuation of the interaction energy $\sigma^2 \equiv \langle (E - \langle E \rangle)^2 \rangle$. The averaging here is performed with respect to all possible realizations of the random density fields $\phi_1(\vec{\rho})$ and $\phi_2(\vec{\rho})$ (see *e.g.*, [3]):

**Eq. 3**
$$P_i[\{\phi_i(\vec{\rho})\}] = C e^{-\int \phi_i^2(\vec{\rho}) d^2\vec{\rho}/2\phi_0},$$

where the normalization constant $C$ is determined from the condition $\int D\phi P[\phi] = 1$. Note that the average magnitude of the local density fluctuations of particles within the area element $\Delta A$, $<\phi_i^2(\vec{\rho})> \Delta A = \phi_0$, is entirely determined by the average particle density $\phi_0$. Therefore, $\phi_0$ is a measure of the fluctuations of the pattern density.

The average energy is easily found in the Fourier representation by denoting

**Eq. 4**
$$\phi_i(\vec{\rho}) = \frac{1}{A} \sum_{\vec{q}} \hat{\phi}_i(\vec{q}) e^{i\vec{q}\vec{\rho}},$$

where $\vec{q}$ is the 2D wave-vector. It is straightforward to show that the averages $<E_0> = E_0$ and $<E_1> = 0$. The quadratic in $\phi_i(\vec{\rho})$ term, $E_2$, has the following form in Fourier representation:

**Eq. 5**
$$E_2 = \frac{1}{A} \sum_{\vec{q}} \hat{\phi}_1(\vec{q}) \hat{\phi}_2(-\vec{q}) \hat{U}(\vec{q}, h),$$

where

**Eq. 6**
$$\hat{U}(\vec{q}, h) = \int U(\vec{r}) e^{i\vec{q}\vec{\rho}} d^2\vec{\rho}.$$

It is obvious now that the average energy is also vanishing, $<E_2> = 0$, as $\hat{\phi}_1(\vec{q})$, and $\hat{\phi}_2(\vec{q})$ are independent variables in the averaging procedure. It is also easy to see that $<E_1^2> = 0$, and $<E_1 E_2> = 0$. The resulting fluctuation $\sigma^2$ of the total energy $E$ is thus determined by only one term, $<|E_2|^2>$:

**Eq. 7**
$$<|E_2|^2> = \frac{1}{A^2} \sum_{\vec{q}} <|\hat{\phi}_1(\vec{q})|^2> <|\hat{\phi}_2(\vec{q})|^2> |\hat{U}(\vec{q}, h)|^2.$$

Performing the Gaussian integration in the Fourier space, $<|\hat{\phi}_i(\vec{q})|^2> = 2\phi_0 A$, we finally obtain

**Eq. 8**
$$\sigma^2_{hetero} = <|E_2|^2> = 4\phi_0^2 A \int |\hat{U}(\vec{q},h)|^2 \frac{d^2\vec{q}}{(2\pi)^2},$$

where we switched from the Fourier sum to the integral. Thus, the probability distribution $P^{(hetero)}(E)$ is the normal distribution with mean $E_0$ and dispersion given by Eq (8).

In the case of random homodimers, the particle density pattern of the second surface is the mirror image of the first one, so

**Eq. 9**
$$E_2 = \int \phi(\vec{\rho}_1) U(\vec{r}_1 - \vec{r}_2) \phi(x_2, -y_2) d^2\vec{\rho}_1 d^2\vec{\rho}_2,$$

The reflection is performed with respect to the *x*-axis; the results are invariant with respect to the choice of the reflection axes. Again, the only relevant term for the energy fluctuations is $<|E_2|^2>$, while the average energy $E_0$ is the same as in heterodimers. Due to the reflection symmetry, the quadratic term in the energy of homodimers reads

**Eq. 10**
$$E_2 = \frac{1}{A} \sum_{\vec{q}} \hat{\phi}(\vec{q})^2 \hat{U}(\vec{q},h),$$

that gives straightforwardly $<E_2> = 0$. The energy fluctuation in homodimers is thus

**Eq. 11**
$$\sigma^2_{homo} = \frac{1}{A^2} \sum_{\vec{q}} <|\hat{\phi}(\vec{q})|^4> |\hat{U}(\vec{q},h)|^2,$$

and performing the Gaussian averaging, we obtain

**Eq. 12**
$$\sigma^2_{homo} = 8\phi_0^2 A \int |\hat{U}(\vec{q},h)|^2 \frac{d^2\vec{q}}{(2\pi)^2},$$

The key result here is that the energy fluctuation for random homodimers is twice as large as the corresponding energy fluctuation for random heterodimers:

**Eq. 13**
$$\sigma^2_{homo} / \sigma^2_{hetero} = 2.$$

This property is universal, it is independent of the type of the interaction potential $U$ and of the density of particles.

We can estimate how the strength of the energy fluctuations depends on the characteristic radius of the potential. We shall choose $U(r)$ to have a Gaussian form, $U(r) = U_0 \exp(-r^2/\xi^2)$, where $\xi$ is the characteristic length of the potential, and $r^2 = \rho^2 + h^2$. The larger is $\xi$, the longer is the range of the potential and the stronger thus are the correlations between the particles. Practically, the characteristic length of the potential, $\xi$, is restricted from below by the size of the particle, $d_0$ (the hard-core size). The most interesting case corresponds to the situation when $\xi \gg h$, this limit corresponds to the strongest effect, when each particle on one surface can make many contacts with particles on the other surface. Performing the Fourier transform of the potential, and taking this limit of a long-range potential, we obtain that for both random homo- and heterodimers the fluctuation of the energy scales as:

**Eq. 14** $$\sigma^2 \sim U_0^2 \phi_0^2 \xi^2 A.$$

The magnitude of the fluctuation is determined by the amplitude of the inter-particle potential and by its characteristic length. The fluctuation is proportional to the total surface area, and to the square of the average density of particles constituting random patterns.

Knowing the distributions $P^{(homo)}(E)$ and $P^{(hetero)}(E)$, one can find the EVDs $P_{min}^{(homo)}(E)$ and $P_{min}^{(hetero)}(E)$ and calculate the average values of the minimum energy. The average of the smallest of $M \sim 2\pi R/\xi$, ($M \gg 1$) values taken from a normal distribution with zero mean and dispersion $\sigma$ is $\langle E_{min} \rangle \approx -\sigma\sqrt{2 \log M}$ [6]. Note that in homodimers, the dispersion of $P^{(homo)}(E)$ is twice as large as in heterodimers, but due to the mirror symmetry there are only $M/2$ independent orientations of the patterns. Thus, one has to consider two EVDs, an EVD of $M$ samples taken from a narrow normal distribution with dispersion $\sigma^2_{hetero}$ (heterodimers), and an EVD of $M/2$ samples from a wider normal distribution with dispersion $2\sigma^2_{hetero}$ (homodimers) . The difference in average lowest energy is thus

Eq. 15 $$\Delta E \sim \sigma_{hetero}\sqrt{\log M} - \sqrt{2}\sigma_{hetero}\sqrt{\log(M/2)}.$$

For large $M$, the prefactor of $\sqrt{2}$ wins over the slower-growing logarithm, so on average, the native state of homodimers has a lower energy than that of heterodimers. The absolute value of this difference is proportional to $\sigma_{hetero} \sim |U_0|\phi_0\xi\sqrt{A}$. The larger is the amplitude of the potential and its correlation length, and the density of patterns, the stronger is the effect. Formally, heterodimers would have a lower energy at $M<8$, however this case is never realized in the disk model. Thus, we argue that the enhanced self-attraction of random patterns, first reported in [2], is a universal effect, at least for sufficiently large surfaces.

Intuitively. the fact that has $P^{(homo)}(E)$ has a larger dispersion compared to heterodimers implies that the corresponding EVD for homodimers will be shifted towards lower energies as compared with heterodimers. Indeed, the EVD is obtained from the low energy tail of $P(E)$, and this tail is shifted towards higher probabilities for homodimers as compared with heterodimers. This is illustrated in Fig. 2, where the energy distributions $P^{(homo)}(E)$, $P^{(hetero)}(E)$ and $P^{(homo)}_{min}(E)$, $P^{(hetero)}_{min}(E)$ are presented . In computing this figure we have assumed that particles interact via a square-well potential, $U(r) = U_0$, if $5\,\text{Å} < r \leq 8\,\text{Å}$, and $U(r) = 0$, if $r > 8\,\text{Å}$, with $U_0 = -2k_BT$, and particles were represented by impenetrable hard spheres of diameter $d_0 = 5\,\text{Å}$. The characteristic length $\xi$ is therefore $\xi = 8\,\text{Å}$. The computed ratio $\sigma_{homo}/\sigma_{hetero} \simeq 1.412$ is very close to its predicted value of $\sqrt{2}$. We have also verified the theoretical prediction of the linear dependence of $\sigma$ on the characteristic length of the potential $\xi$ (inset in Fig. 2). The computed ratio of the linear fit slopes, $\sigma_{homo}/\sigma_{hetero} \simeq 1.416$, is again in excellent agreement with the predicted value of $\sqrt{2}$. The deviation from the linear behavior of $\sigma$ at small values of $\xi$ (short-range potential, when $\xi$ is very close to $d_0$ and $h$) is due to the fact that there are very few contacts between the particles across the interface possible, and besides $\xi$, the two additional length-scales, $d_0$ and $h$, become significant.

In summary, we confirmed theoretically the prediction [2] of universally enhanced self-attraction of random patterns. We predicted here that the magnitude of the energy fluctuations for homodimers is always twice as large as compared with one for heterodimers. This exact result holds true for any type of the inter-particle interaction potential, and for random patterns with arbitrary number of types of interacting particles. This implies that the distribution of lowest energies in pairs of interacting surfaces (the EVD) is always shifted towards the lower energies for homodimers as compared with heterodimers, in agreement with the computational prediction [2]. The effect stems from two principal sources, the mirror symmetry ensuring the difference in dispersion of energy distributions, and the slow ($\sim \sqrt{\log M}$) dependence of the mean value of the EVD on the number of samples $M$. Our results may be relevant for interpretation of two important experimental observations, the anomalously high frequency of homodimers in protein interaction networks of different organisms [4], and the enhanced propensity to aggregate found in proteins with similar aminoacid sequences [5]. We suggest that both of these phenomena might be an evolutionary manifestation of the general physical principle of enhanced self-attraction predicted in our work.

**FIGURE CAPTIONS**

Figure 1: Snapshot of a random pattern.

Figure 2: Computed probability distribution $P(E)$, and EVD, $P_{\min}(E)$, for heterodimers and homodimers, respectively. We generated $10^6$ surfaces with random patterns, where each surface has the diameter, $D = 140 \text{ Å}$. We placed $N_0 = 350$ particles (at random) on each surface, with the hard-core diameter of a particle being $d_0 = 5 \text{ Å}$, (and the average surface fraction of each pattern is thus $N_0 d_0^2 / D^2 \simeq 0.446$. The potential $U(r)$ was chosen to be a square-well with the amplitude, $U_0 = -2k_B T$, where $k_B$ is the Boltzmann constant and $T$ is the temperature, and with the length, $\xi = 8 \text{ Å}$ (i.e. $U(r) = U_0$, if $5 \text{ Å} < r \leq 8 \text{ Å}$, and $U(r) = 0$, if $r > 8 \text{ Å}$), $E$ is plotted in the units of $k_B T$, and normalized by the total number of interface particles. $P(E)$ is normalized in such a way that $\int P(E) \, dE = 1$. The inter-surface separation, $h$, was chosen to be, $h = 5.01 \text{ Å}$, i.e. such that the surfaces are practically in contact. Inset: Computed dependence of $\sigma$ as a function of $\xi$ for heterodimers and homodimers, respectively. Straight lines represent the linear fits to the data. The linear correlation coefficient is $R \simeq 0.99$ in both cases. The computed ratio of the fits' slopes, $\sigma_{\text{homo}} / \sigma_{\text{hetero}} \simeq 1.416$, is in excellent agreement with the theoretical prediction, $\sqrt{2} \simeq 1.414$.

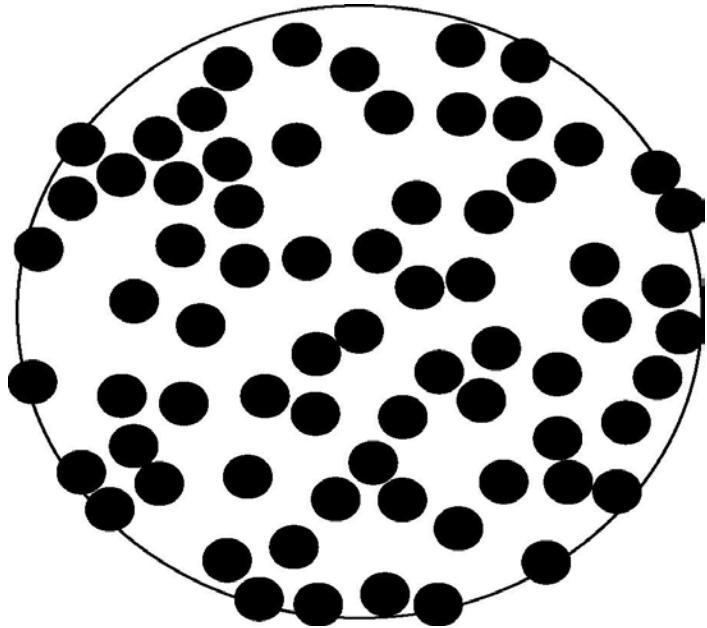

Figure 1

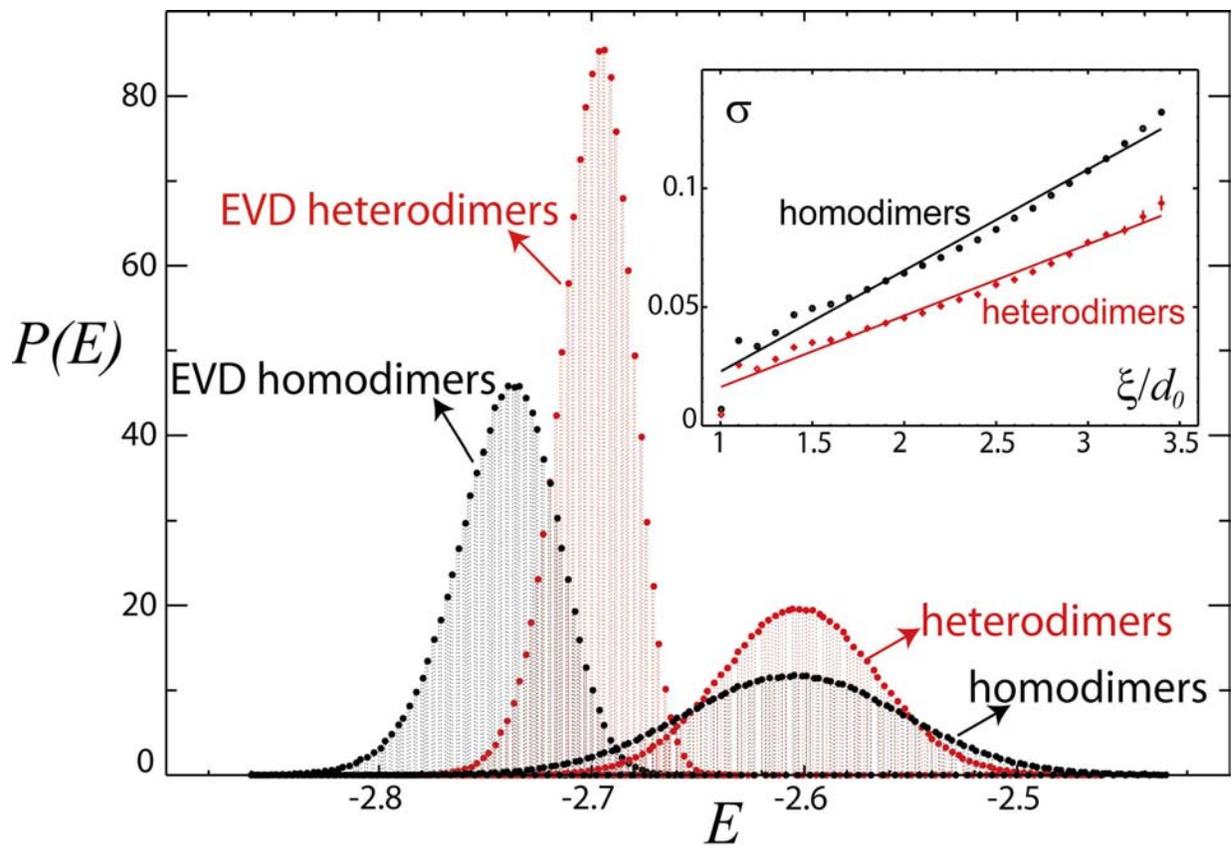

Figure 2